\documentclass[12pt,preprint]{aastex}

\def\ltsima{$\; \buildrel < \over \sim \;$}
\def\gtsima{$\; \buildrel > \over \sim \;$}
\def\lsim{\lower.5ex\hbox{\ltsima}}
\def\gsim{\lower.5ex\hbox{\gtsima}}
\def\lapp{\ifmmode\stackrel{<}{_{\sim}}\else$\stackrel{<}{_{\sim}}$\fi}
\def\gapp{\ifmmode\stackrel{>}{_{\sim}}\else$\stackrel{<}{_{\sim}}$\fi}

\def\media#1{\langle #1\rangle}

\newdimen\minuswidth    
\setbox0=\hbox{$-$}
\minuswidth=\wd0
\catcode`@=\active
\def@{\kern\minuswidth}
\setbox0=\hbox{\rm0}
 
\shorttitle{Surface Density Profile of NGC6388}
\shortauthors{Lanzoni et al.}
 
\begin{document} 
\title{The surface density profile of 
NGC~6388:  a good candidate for harboring an intermediate-mass black hole
\footnote{Based on observations with the NASA/ESA {\it HST}, obtained at the
Space Telescope Science Institute, which is operated by AURA, Inc., under
NASA contract NAS5-26555. Also based on WFI observations collected
at the European Southern Observatory, La Silla, Chile.} }

\author{
B. Lanzoni\altaffilmark{1},
E. Dalessandro\altaffilmark{1,2},
F.R. Ferraro\altaffilmark{1},
P. Miocchi\altaffilmark{3},
E. Valenti\altaffilmark{4},
R.T. Rood\altaffilmark{5}
}
\affil{\altaffilmark{1} Dipartimento di Astronomia, Universit\`a degli Studi
di Bologna, via Ranzani 1, I--40127 Bologna, Italy}
\affil{\altaffilmark{2} ASI,Centro di Geodesia Spaziale, contrada Terlecchia,
I-75100, Matera, Italy }   
\affil{\altaffilmark{3} Dipartimento di Fisica, Universit\`a di Roma ``La
Sapienza,'' P.le A. Moro 5, I--00185 Roma, Italy}
\affil{\altaffilmark{4} European Southern Observatory,
 Alonso de Cordova 3107, Vitacura, Santiago, Chile}
\affil{\altaffilmark{5} Astronomy Department, University of Virginia,
P.O. Box 400325, Charlottesville, VA, 22904}
\date{24 July, 07}

\begin{abstract}
We have used a combination of high resolution ({\it HST} ACS-HRC,
ACS-WFC, and WFPC2) and wide-field ({\it ESO}-WFI) observations of the
galactic globular cluster NGC~6388 to derive its center of gravity,
projected density profile, and central surface brightness profile.
While the overall projected profiles are well fit by a King model
with intermediate concentration ($c=1.8$) and sizable core radius
($r_c=7\farcs2$), a significant power law  (with slope
$\alpha=-0.2$) deviation from a flat core behavior has been detected
within the inner $1\arcsec$.  These properties suggest the
presence of a central intermediate mass black hole.  The observed
profiles are well reproduced by a multi-mass isotropic,
spherical model including a black hole with a mass of $\sim 5.7\times
10^3\, M_\odot$.
\end{abstract}
 

\section{INTRODUCTION}
The surface brightness (SB) and the projected density profiles of the
vast majority of globular clusters (GCs) are well reproduced by a
family of simple models characterized by an extended isothermal core
and a tidally trucated envelope---the so-called King models (King
1966).  However a fraction ($\sim 15$--20\%, see  \citet{dk86}) 
of galactic GCs deviate significantly from this behavior. The
projected density profiles of these clusters do not exhibit an
extended core, showing instead a power law behavior $\Sigma(r)\propto
r^\alpha$ with $\alpha$ ranging from $-0.8$ to $-1.0$ .  This feature
has been thought to arise from the dynamical evolution of stellar
systems that have experienced the collapse of the core. These are
called post-core collapse clusters, hereafter PCC).

However, other processes can affect the shape of a star cluster density
profile; among these the existence of an Intermediate Mass Black Hole (IMBH)
in the central region has recently received attention. Interestingly enough,
detailed collisional $N$-body simulations (Baumgardt et al., 2005, hereafter
BMH05; Trenti et al. 2007) and theoretical arguments \citep{heggie07} have
shown that the presence of a IMBH yields quite a different SB profile than
core collapse does.  These studies indicate that in an initially dense
cluster a IMBH gives rise to a strong expansion of the central region that,
in turn, leads to a quasi-steady configuration resembling that of a medium
concentration cluster with a core-like profile.  Thus, the clusters most
likely to harbor IMBHs are those having the appearance of normal King model
clusters except in the very central regions where a power law deviation from
a flat behavior is expected.  The exponent of this power law is predicted to
be significantly lower ($\alpha\sim -0.2$) than in the PCC case (BMH05,
Miocchi 2007). Small departures from a King model have been
observed in the projected density profile of a 
few GCs (included NGC6388) by Noyola \& Gebhardt (2006, hereafter
NG06).

The confirmation of the existence of IMBHs and an estimate of their
frequency would be important for a number of astrophysical problems
like the formation processes of super massive BH in galaxies,
super-Eddington X-ray sources in extragalactic globular clusters
\citep{sivakoff07}, the origin of ultraluminous X-ray sources
\citep{miller03,fa06}.  Despite of their potential importance, the
existence of IMBH in GCs is still an open question.  For instance, the
evidence for an IMBH in M15 reported by \citet{vdm02} and
\citet{ger02} has been questioned by \cite{bau03a}.  \citet{bau03b}
also question evidence for an IMBH in G1 in M31 \citep{ge02} (but see
\citet{ge05} and the recent findings by Ulvestad et al (2007) 
and Green \& Ho, 2007).

Here we present accurate surface density and SB profiles obtained with
a combination of high-resolution and wide-field observations of the
galactic globular cluster NGC~6388, which a number of authors
(BMH05; NG06; Drukier \&
Bailyn 2003, hereafter DB03; Miocchi 2007) have 
indicated as a prime candidate to
harbor an IMBH.  These profiles nicely match the
theoretical models of a cluster with a $5.7\times 10^3
M_\odot$ BH: having an extended core,
intermediate concentration, but also significant deviations from a flat
core distribution in the innermost cluster regions ($r\lsim
1\arcsec$).  

\section{THE DATA}
In this paper we make use of a combination of high-resolution and
wide-field photometric data sets, obtained with WFPC2 and ACS on board
{\it HST}, and with WFI at ESO-2.2\,m, respectively.  A detailed
description of the observations, photometric reduction and astrometry
procedures of the data obtained with WFPC2, ACS-WFC, and WFI is given
in a companion paper discussing the Blue Straggler Star and Horizontal
Branch populations \citep{ema07}.  Here we use only the optical ($B$,
$V$, $I$) samples from the entire multi-wavelength data set. These
have been all homogenized and transformed to the Johnson magnitude
system. All the star positions have been placed on the absolute
astrometric system using several hundred astrometric reference
stars from the new astrometric 2-MASS catalogue\footnote{Available at
{\tt http://irsa.ipac.caltech.edu}.},
following the procedure described, e.g., in \citet{lan07},  
with a final astrometric accuracy of the order of $\sim 0\farcs3$ both in RA
and Dec.

\begin{figure}[!hp]
\begin{center}
\includegraphics[scale=0.7]{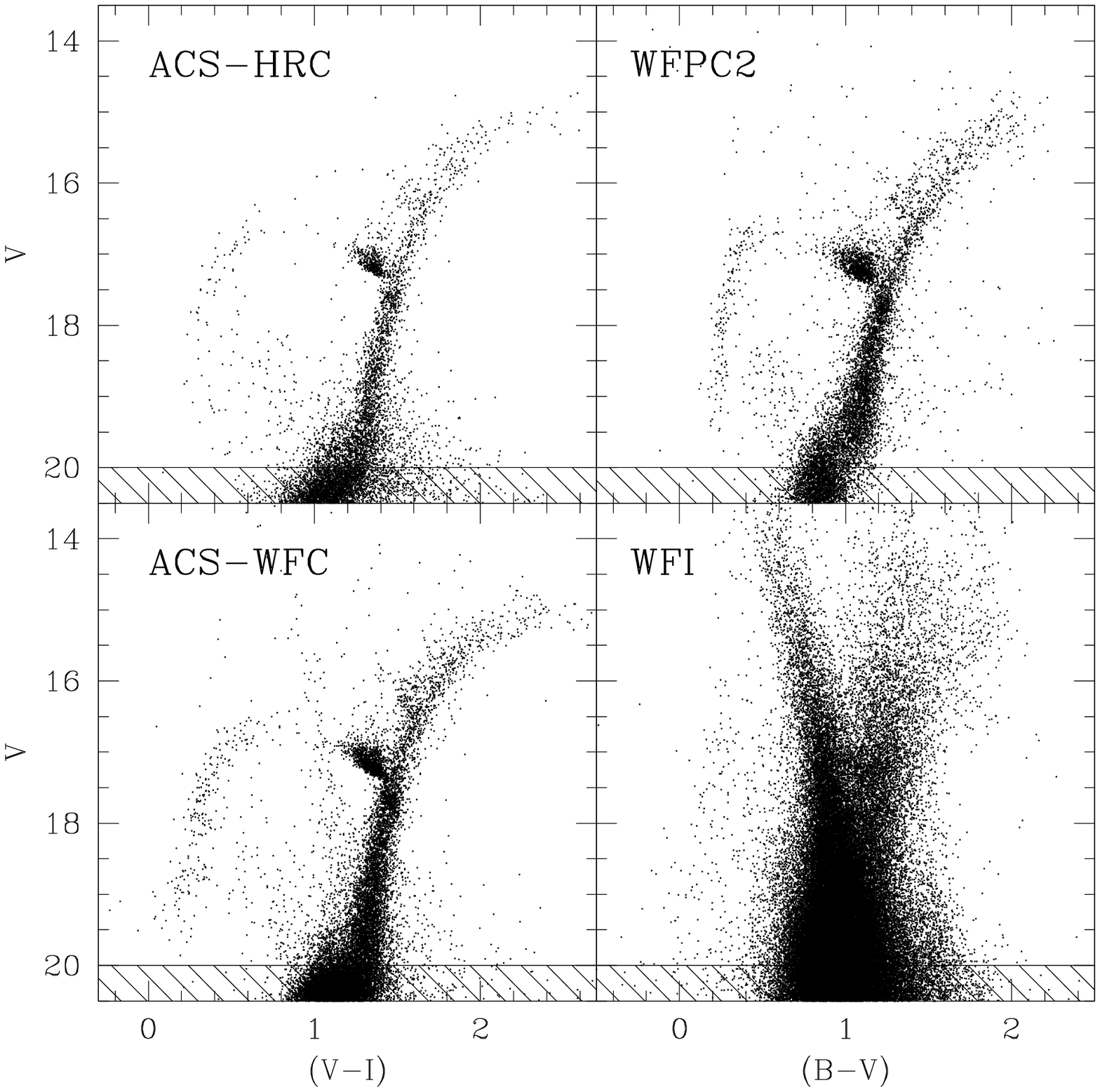}
\caption{Color Magnitude diagrams for the four data sets used. The WFI
data are dominated by field contamination.}
\label{fig:cmd}
\end{center}
\end{figure}

Additional images obtained with the ACS-HRC have been retrieved from the
ESO/ST-ECF Science Archive.
These data sample the cluster central region with a field of view
(FoV) of $26\arcsec\times 29\arcsec$ and a spatial resolution of
$0\farcs027\,{\rm pix}^{-1}$.  The HRC data were obtained through
filters $F555W$ ($V$) and $F814W$ ($I$), with total exposure times of
620 and 3070\,s, respectively.  After corrections for geometric
distortions and effective flux \citep{si05}, the photometric
analysis was performed
by using SExtractor \citep{be96}, adopting a fixed aperture radius of
4 pixels ($0\farcs108$).  The sample has then been astrometrized and
photometrically calibrated by using the stars in common with the
ACS-WFC catalog. The color-magnitude diagrams based on the data from
all four data sets are shown in Fig.~\ref{fig:cmd}.

\section{CENTER OF GRAVITY}
\label{sec:cgrav}
Given the absolute positions of individual stars in each catalog, the
center of gravity, $C_{\rm grav}$, of NGC~6388 has been determined by
averaging the coordinates $\alpha$ and $\delta$ of all stars detected
in the highest resolution data set (the HRC sample).  In order to
correct for spurious effects due to incompleteness in the very inner
regions of the cluster, we considered only stars brighter than $V=20$
(roughly corresponding to the  sub-giant branch of the
cluster).  By following the iterative procedure described in
\citet[][see also Ferraro et al. 2004]{mont95}, the center of
gravity is located at $\alpha_{\rm J2000} = 17^{\rm h}\,
36^{\rm m}\, 17\fs23$, $\delta_{J2000} = -44\arcdeg\,44\arcmin\,
7\farcs1$, with an uncertainty of $0\farcs3$ in both $\alpha$ and
$\delta$.  A careful examination of field inside the core radius shows
that our determination of the center is
biased neither by the presence of very bright stars, nor of a star clump.
The derived $C_{\rm grav}$ is located $\sim 2\farcs6$ southeast
($\Delta\alpha = 3\farcs4$, $\Delta\delta=-1\farcs1$) of that
derived by \citet{dm93} using the surface
brightness distribution.
An accurate comparison with the center adopted by NG06 is not 
possible since the value listed in their Table 1 is just referred 
to the world coordinate system of a specific WFPC2 image, however 
a visual inspection suggests that their center is
located $\sim 0\farcs5$ SE of ours.

\section{PROJECTED DENSITY AND SURFACE BRIGHTNESS PROFILES}
\label{sec:prof}
We have determined the projected density profile of NGC~6388 using
direct star counts over the entire cluster radial extent, from $C_{\rm
grav}$ out to $\sim 1400\arcsec\sim23\arcmin$.  This
distance is significantly larger than the expected cluster extension
($r_t=372\arcsec$, Harris 1996). In order to avoid spurious effects
due to possible incompleteness, only stars brighter than $V=20$ have
been considered (see Dalessandro et al 2007). There 
are more than 58,000 stars in the entire (i.e.,
the combination of ACS, WFPC2 and WFI) photometric data set.
Following the procedure described in \citet[][]{fe99a}, we have
divided the sample in 40 concentric annuli, each centered on $C_{\rm
grav}$. Each annulus has been split into an adequate number of sub-sectors.
The number of stars lying within each sub-sector was counted, and the
star density was obtained by dividing these values by the
corresponding sub-sector areas.  The stellar density in each annulus
was then obtained as the average of the sub-sector densities, and the
uncertainty in the average values for each annulus was estimated from
the variance among the sub-sectors. The radius associated with each
annulus is the mid-point of the radial bin.  The outermost ($r\gsim
7\arcmin$) measures have an almost constant value and their average
has been used to estimate the Galaxy contamination to be
$\sim 56$ stars arcmin$^{-2}$. Subtracting this background yields the
the profiles shown in Figs.~\ref{fig:nbcounts}.

If the innermost ($r<1\arcsec$) points are excluded, the density
profile is well fit all over the entire extension by an isotropic,
single-mass King model with a core of $r_c=7\farcs2$ and an
intermediate concentration ($c=1.8$). These values are similar to
those quoted by Trager et al. (1995; $r_c=7\farcs4$ and $c=1.7$),
Harris (1996; $r_c=7\farcs2$ and $c=1.7$), and McLaughling \& van der
Marel (2005; $r_c=7\farcs8$ and $c=1.71$).\footnote{Note that the
higher concentration quoted in the present paper implies a ($\sim
20\%$) larger tidal radius ($r_t=454\arcsec$) than previously
determined.}  In the inner $\sim 1\arcsec$ the observed profile shows
an indication of a deviation from a flat core behavior.  With only 7 stars 
in the innermost bin, the statistical error of the inner bin
($0\arcsec\le r <0\farcs3$) is relatively large. With star counts
alone the exact amount of the deviation from the flat core cannot
be reliably estimated.
 
Exploiting the exceptional high resolution of ACS-HRC images we have
computed the SB profile by direct aperture photometry to more
accurately determine the inner shape of the cluster profile. In the
innermost region ($r<1\arcsec$) we have used two sets of annuli
stepped by $0\farcs3$ and $0\farcs5$, respectively. The SB values have
been computed as the sum of the counts in each pixel, divided by the
number of pixels in any given annulus. The counts have then been
converted to a magnitude scale and calibrated by using a relation that
links the ``instrumental'' magnitude to the calibrated one (obtained
by performing aperture photometry for a number of high S/N isolated
stars).  The resulting SB profile for the innermost $10\arcsec$ from
the center is shown in Figure \ref{fig:mu}.  A steepening of the
profile at $r\lsim 1\arcsec$ is clearly apparent, in agreement with
what we found above for the surface density distribution.  A linear fit
to the inner points suggest that the slope of the profile is
$0.6\pm 0.06$ in the $\log\Sigma-\log r$ plane.  
 In terms of the surface luminosity
density $I(r)$, if $I\propto r^\alpha$ we find $\alpha\simeq -0.23$ .
This is steeper (but still marginally consistent within the errors)
than the slope $\alpha=-0.13\pm 0.07$ derived from the analysis of
WFPC2 images by NG06.  
The N06 profile is  shown for comparison in Figure 3, as can be seen
their profile is fully compatible with our data in the common
region. The use of high resolution images  allow us to probe the 
innermost region of the cluster where most of the deviation from a
flat behavior occurs\footnote{Even the small
difference in the center determination can play a role.
Simulations have shown that even an offset of only a few $0\farcs1$
is sufficient to flatten the profile. An additional difference in the
slope determination  might arise from the different approach
used by NG06, who removed the bright stars and did not fit a power
law to the data but instead calculated the derivative of the smooth 
central profile.}.

\section{DISCUSSION}
\label{sec:disc}
The properties of the projected density and SB profiles derived in the
previous section for NGC~6388 are not those of a cluster which has
experienced the core collapse. Instead they are just what BMH05
suggest as the signatures of a cluster harboring an IMBH in its
center: {\it (i)} a typical King profile with intermediate
concentration ($c=1.8$) in the external regions, {\it (ii)} a sizable
core, and {\it (iii)} a inner logarithmic slope $\alpha\sim -0.2$.
These features have been recently confirmed by the predictions of a
self-consistent parametric model that includes the presence of a
central IMBH \citep{miocchi07}.  This model consists of a multi-mass
isotropic, spherical King model, which has been extended inside the
region of gravitational influence of the BH via the inclusion of the
\citet{BW} phase-space distribution function.

\begin{figure}[!hp]
\begin{center}
\includegraphics[scale=0.7]{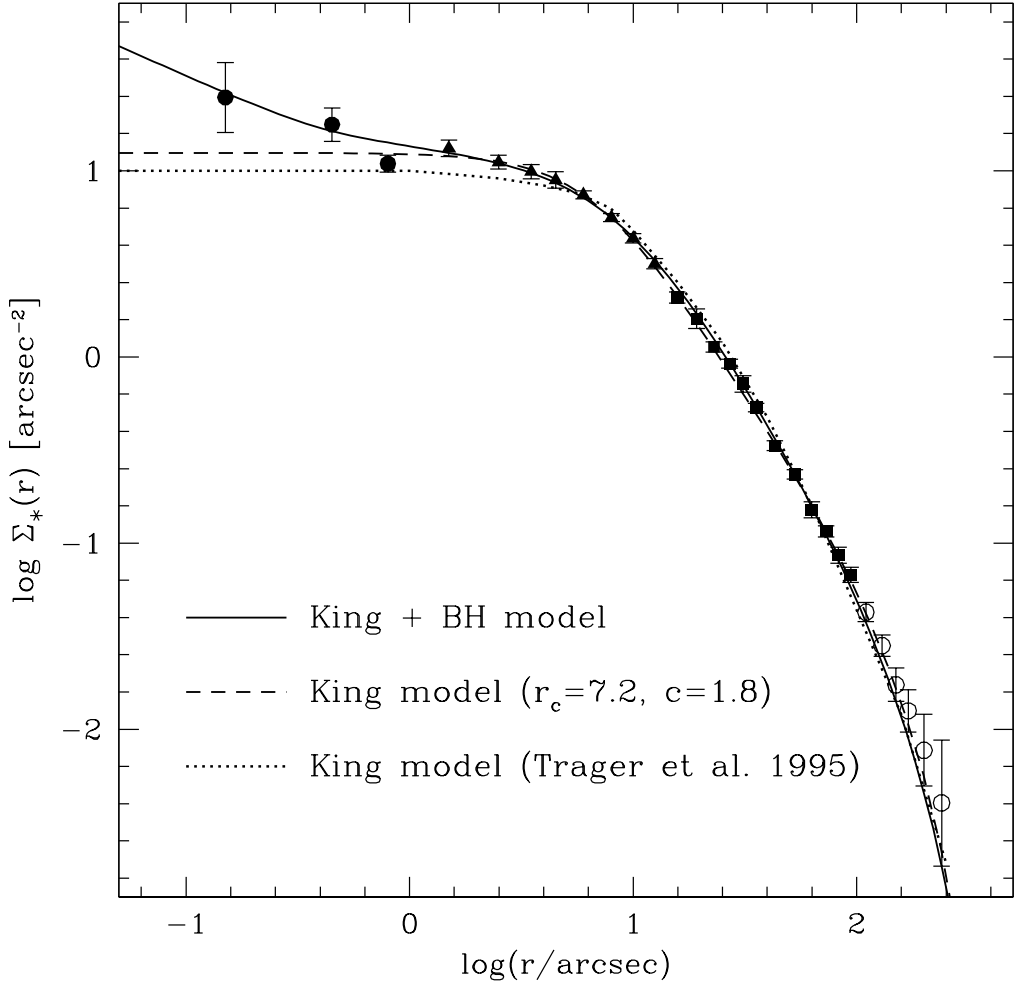}
\caption{Observed surface density profile obtained by star counts from
the combined photometric data-set: ACS-HRC ({\it filled circles}),
WFPC2 ({\it triangles}), ACS-WFC ({\it squares}), and WFI ({\it empty
circles}). The background level (see Sect. \ref{sec:prof}) has been
subtracted. The single-mass King model that best fits the profile
excluding the innermost ($r\lsim 1\arcsec$) points is
shown as a dashed line. It is characterized by a sizable core radius
($r_c=7\farcs2$), and an intermediate concentration ($c=1.8$). The
solid line shows the profile of the model (see Sect. \ref{sec:disc})
including a $5.7\times 10^3 M_\odot$ BH in the cluster center.
The profile from Trager et al (1995) is shown ({\it dotted line}
for comparison. 
}
\label{fig:nbcounts}
\end{center}
\end{figure}

\begin{center}
\begin{figure}[!p]
\includegraphics[scale=0.7]{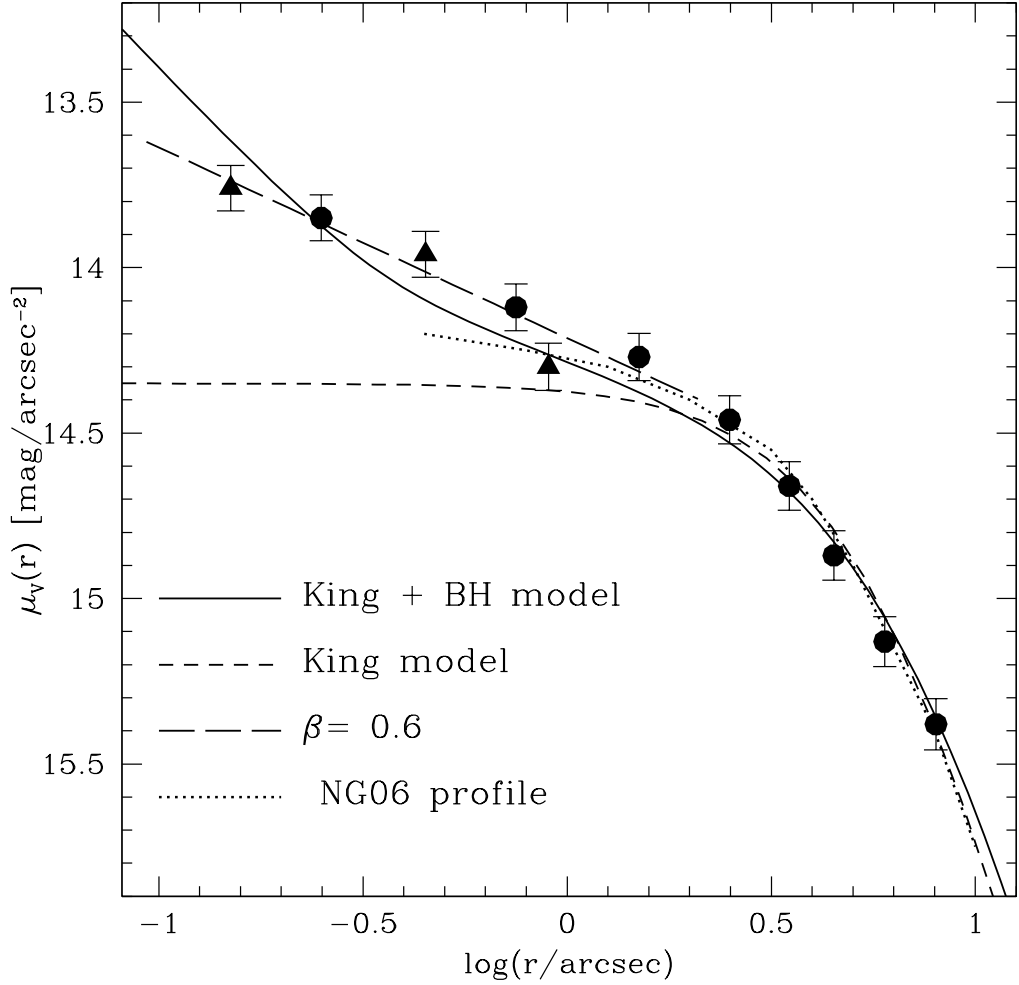}
\caption{Surface brightness profile derived from the ACS-HRC images
within $10\arcsec$ from the cluster center. {\it Dots} refer to a radial
binning of $0\farcs5$, triangles sample the inner $1\arcsec$ by
steps of $0\farcs3$. The {\it solid line} shows the profile of the model
obtained by including a $5.7\times 10^3 M_\odot$ BH (the same as in
Fig. \ref{fig:nbcounts}). The {\it short-dashed line} corresponds to the King
model shown in Fig. \ref{fig:nbcounts}.
$\beta =0.6$ is the slope of the linear best fit 
(see the {\it long-dashed line}) to the innermost points. 
The profile obtained by NG06 is also shown (see the {\it dotted line} for comparison.}
\label{fig:mu}
\end{figure}
\end{center}

In order to further support the case for an IMBH in the center of
NGC~6388, we have used this model to reproduce the observed density
and SB profiles.  A Salpeter mass function ($\mathrm{d}N\propto
m^{-1.35}\mathrm{d}\log m$) is assumed in the model and seven discrete
mass bins are used to approximate the continuum mass spectrum of the
real cluster. The stellar mass range in the interval from 0.3 to
0.9\,$M_\odot$ (here and thereafter, where not specified, masses are
measured in $M_\odot$) equally subdivided in 6 bins $0.1$ wide. These
were populated with main sequence stars.  In addition the bins
$[0.5,0.6]$ and $[0.7,0.8]$ include WD populations with mass $0.55$
and $0.75$, coming from progenitors with mass, respectively, in
$[0.9,1.5]$ and $[1.5,4]$ ranges. The seventh mass bin
contains a massive WD population with $m=1.2$, hypothosized as
remnants of stars with mass 4--8. The $[0.8,0.9]$ bin has
$\media{m}=0.84$ and contains the TO stars, plus giants and HB
stars. The light-to-mass ratios were taken to be $\{4.9\times
10^{-3},10^{-2}, 2.3\times 10^{-2},6.5\times 10^{-2},0.19,10,0\}$
corresponding to the bins ordered in increasing average mass.  The
velocity dispersion of the seven components is constrained by the
requirement of complete energy equipartition at the border of the BH
influence region \citep[see][~for details]{miocchi06,miocchi07}, where
the adimensionalised potential $W_\mathrm{BH}$, along with the  
the ratio between the BH mass ($M_\mathrm{BH}$) and 
the cluster mass ($M$),
 determine the form of the
generated profiles. Besides of the various scale parameters,
$W_\mathrm{BH}$ and $M_\mathrm{BH}$ are adjusted to obtain the best fit to
the observed profiles. To do this, we 
conservatively include only data from the central $100\arcsec$
in order to  avoid possible spurious
effects which might affect the most external points of the SB profiles 
because of the field contamination
subtraction.
The best fit  to the observed SB  profile is then
found for $r_c=7\farcs2$, $c=1.8$, $W_\mathrm{BH}=11.5$
 and
$M_\mathrm{BH}=2.2\times 10^{-3}M$
(yielding $P(\chi^2>\chi^2_\mathrm{fit})>99$\%). 
The level of confidence remains above 97\% for an IMBH
mass in the range $2.1$--$2.4 \times 10^{-3}M$.
The $r_c$ and $c$ values  are consistent with
the value deduced by the parametric fit of the single-mass King model
mentioned above.
The results of this parametric fit to the projected density and the SB
profiles are shown in Figs. \ref{fig:nbcounts} and \ref{fig:mu}
respectively.

 By assuming the total cluster luminosity $V_t=6.72$
\citep{har96}, and the distance modulus $(m-M)_V=16.59$
\citep{ema07}\footnote{The distance has been obtained differentially
with respect to 47 Tucanae, by assuming the distance scale by Ferraro
et al. (1999b).}, we estimate a total mass of $2.6\times 10^6 M_\odot$
for NGC~6388, yielding $M_{\rm BH}\simeq 5.7\times 10^3
M_\odot$.\footnote{Note that this mass is well within the range of
values (2.5--$10 \times 10^3$) derived from the $M_{\rm BH}-\sigma$
relation (Ferrarese \& Merritt 2000; Gebhardt et al. 2000; see
for example equation (10) of DB03), by assuming the low
($\beta=4.02$) and the high ($\beta=4.65$) exponents, respectively.}

While the central IMBH is a possible explanation of the shape of the
observed profiles, one might question whether this result is
unique. In fact, a central concentration of massive remnants (like
white dwarfs and neutron stars) has been proposed as an alternative
explanation in the case of M15 \citep{vdb06}. However,
we have found that a multi-mass King model including a population of
such remnants but without central BH is unable to reproduce the
observed slope of the profiles in the core region.
Since our
evidence is based
exclusively from the shape of the density profile, the presence of a
IMBH at the center of NGC~6388 is still debatable. Accurate
kinematical studies of the motion of stars in the central region of
the cluster are needed to solidify the case.

The region in which to seek for the possible kinematical signatures of
a BH is very small.  The self-consistent model here employed generates
a projected velocity dispersion profile that shows a sharp rise from
the ``isothermal plateau'' to a purely Keplerian behavior at $r\sim
0\farcs16$ (i.e., $\sim 0.02\, r_c$).  A more promising path to detect
the kinematic signature of a BH is by proper motion measurements
(DB03). Some stars should move with anomalously high
velocities under the influence of the BH.  In order to estimate the
number of these stars, we first need to evaluate the BH radius of
influence $r_h$.  A crude estimate of $r_h$ is given by: $r_h =
G\,M_{\rm BH}/\sigma_0^2$, where $\sigma_0$ is the velocity dispersion
outside $r_h$. By assuming $\sigma_0=18.9\,{\rm km\,s^{-1}}$
\citep{pry93}, we find $r_h \simeq 0.07$\,pc, corresponding to
$1\farcs1$ with the cluster distance \citep[$d=13.18$\,kpc;][]{ema07}.
By using equation (7) of DB03, it is possible to
estimate the number of stars ($N$), measurable through proper motions
studies, having velocities 3 times the cluster velocity dispersion
outside $r_h$: $N=0.27\, \Sigma_0\, r_h^2$, where $\Sigma_0\, r_h^2$
is the number of stars within $r_h$. This relation suggests that $\sim
27\%$ of stars within $r_h$ are expected to show anomalously high
velocity.  We can directly derive this number from the HRC
images. Adopting the value of the cluster center presented above we
count 28 relatively bright ($V<19$) stars within $r_h=1\farcs1$
(out of a total of $\sim 85$ stars detected down to $V\sim 22$),
corresponding to a total of $\sim 7$ high-velocity stars. This
estimate shows that the presence of $\sim 5\times 10^3 M_\odot$ BH in
the center of NGC~6388 can in principle be kinematically confirmed in
the near future through accurate proper motion studies  
or radial velocity measurements with {\it Adaptive Optic}
supported instruments.  However,
these measurements can be quite challenging. According to
Figure 1 of DB03, the  high velocity stars are expected to be mainly
confined within 0.4\,$r_h$ (only $\sim 0\farcs4$ from
the center). They would have speeds of order
$60\,{\rm km\,s^{-1}}$ which, given projection effects, would give
tangental velocities of order $20\,{\rm km\,s^{-1}}$. The distance
derived by \citet{ema07} and adopted here is 25\% larger than that
from Harris (1996), so the resulting proper motions would be $\sim
0.3\,{\rm mas\,yr^{-1}}$. Given  the current estimate for the
proper motion  measurement error of roughly 0.3\,mas,  
a baseline of at least
3--5\,yr  would be required.

\acknowledgements This research was supported by contract ASI-INAF
I/023/05/0, PRIN-INAF2006 and by the Ministero dell'Istruzione,
dell'Universit\'a e della Ricerca.  ED is supported by ASI.  RTR is
partially supported by STScI grant GO-10524.  This research has made
use of the ESO/ST-ECF Science Archive facility which is a joint
collaboration of the European Southern Observatory and the Space
Telescope - European Coordinating Facility.

\clearpage 

\end{document}